\documentclass[aps,prb,
 amsmath,amssymb,
 reprint,
]{revtex4-2}

\usepackage{graphicx}
\usepackage{dcolumn}
\usepackage{bm}
\usepackage{textgreek}
\usepackage[utf8]{inputenc}
\usepackage[LGR,T1]{fontenc}
\usepackage{etoolbox}
\usepackage{float}
\usepackage{hyperref}
\hypersetup{colorlinks=true,linkcolor=blue,filecolor=gray,urlcolor=blue,citecolor=blue}
\newcommand{\angstrom}{\textup{\AA}}
\DeclareUnicodeCharacter{2212}{\textendash}

\begin{document}

\title{Phase transitions of correlated systems from graph neural networks with quantum embedding techniques}
\author{Rishi Rao}

\author{Li Zhu}
\email{li.zhu@rutgers.edu}
\affiliation{ 
Department of Physics, Rutgers University, Newark, New Jersey 07102, USA
}%

\date{\today}

\begin{abstract}  

     Correlated systems represent a class of materials that are difficult to describe through traditional electronic structure methods. The computational cost of simulating the structural dynamics of such systems, with correlation effects considered, is substantial. Here, we investigate the structural dynamics of $f$- and $d$-electron correlated systems by integrating quantum embedding techniques with interatomic potentials derived from graph neural networks. For Cerium, a prototypical correlated $f$-electron system, we use Density Functional Theory with the Gutzwiller approximation to generate training data due to efficiency with which correlations effects are included for large multi-orbital systems. For Nickel Oxide, a prototypical correlated $d$-electron system, advancements in computational capabilities now permit the use of full Dynamical Mean Field Theory to obtain energies and forces. We train neural networks on this data to create a model of the potential energy surface, enabling rapid and effective exploration of structural dynamics. Utilizing these potentials, we delineate transition pathways between the $\alpha$, $\alpha'$, and $\alpha''$ phases of Cerium and predict the melting curve of Nickel Oxide. Our results demonstrate the potential of machine learning potentials to accelerate the study of strongly correlated systems, offering a scalable approach to explore and understand the complex physics governing these materials. [This is a preprint of Phys. Rev. B 110, 245111 (2024)]

\end{abstract}

\maketitle

\section{Introduction}

Strongly correlated systems, characterized by significant electron-electron interactions, present a frontier in materials science and condensed matter physics. These interactions lead to phenomena like Mott transitions \cite{Mott1970,Belitz1994}, heavy fermion behavior \cite{Stewart1984, Si2010,raoPredictingNewHeavy2023a}, spin-charge separation \cite{Zakrzewski2018,Evans2008} and other correlation induced effects that can be technologically useful and physically interesting \cite{Cleland2020,Sugata2003,Molinari2019,Timoshenko2017,Yamaguchi1987}. 
Conventional computational methods for dealing with interactions, such as Dynamical Mean Field Theory (DMFT) \cite{Arpita2019, Antoine1996}, Density Matrix Renormalization Groups (DMRG) \cite{Schollwock2005,Baiardi2020}, Gutzwiller wavefunction techniques \cite{gutzwillerJulien2005,Deng2009,Ho2008}, and Hubbard U corrections to Density Functional Theory (DFT) \cite{Himmetoglu2014,Albers2009}, have advanced our understanding of the electronic structures of such materials. However, simulating the structural dynamics and thermodynamics of strongly correlated materials remains a significant challenge due to the computational cost associated with the large Hilbert spaces that arise from many-body electron-electron interactions \cite{Adler2019,Pavarini2021}. As the number of orbitals increase, the Hilbert space of commonly used models, such as the multi-orbital Hubbard model or the Anderson impurity model, rapidly increases as well. This expansion makes the calculation of dynamics particularly tedious for correlated systems, as dynamics generally require extensive sampling over many structural configurations.

This computational bottleneck becomes particularly prohibitive when exploring structural phase transitions and conducting molecular dynamics (MD) simulations. Although methods like DMFT and DMRG can offer nearly quantitative accuracy, their scalability issues and the technical difficulties in applying these approaches to multi-dimensional and large multi-orbital systems limit their practicality for direct studies of structural dynamics. For example, the exact method for solving the DMFT impurity problem, the continuous time quantum monte carlo (CTQMC) algorithm \cite{Werner2006,Turkowski2021,Gull2011}, samples Feynman diagrams to arbitrary order in imaginary time. This suffers from a large number of Monte Carlo steps required to accurately gauge the self energy as temperature is lowered or when considering a greater number of orbitals.
Similarly, DMRG becomes technically difficult to implement in more than two dimensions \cite{Yanai2015}. 
Techniques like the Hubbard U correction to DFT Hamiltonians (DFT+U), while useful, often lack the quantitative accuracy needed for predicting complex behaviors under varying conditions such as high pressure \cite{Panda2017}, and are less quantitatively accurate than DMFT \cite{Amadon2012,Hariki2017}. 
The Gutzwiller approximation technique, equivalent to a mean field approximation of slave boson techniques in the limit of infinite spatial dimensions \cite{Metzner1989,Bunemann2007,Lanata2017}, has been successfully applied to numerous correlated systems \cite{Wang2010,Tian2015,Tian2011,Lu2013,Schickling2012,Zhou2010,Lanata2013} and provides a relatively cheap way method for tackling large multi-orbital systems. However, even this method is significantly more computationally expensive than traditional DFT as it provides iterative corrections to a tight-binding Hamiltonian generated from DFT, and is primarily a zero-temperature technique.

These computational challenges are significant in studying materials like Cerium (Ce) and Nickel Oxide (NiO), which exhibit interesting properties under extreme conditions. Cerium, known for its complex structural phase transitions under intermediate pressure, exemplifies the challenges associated with accurately determining the lowest energy phase under varying pressure and temperature conditions. These phases are critical for understanding its properties and potential applications, yet first-principles study is hampered by the prohibitive computational cost of simulating strong correlation effects. Similarly, determining the melting point of compounds like NiO, an important component of the lower mantle of Earth \cite{nioHerzberg2013,nioMcDonough1995,nioWalter1998}, under high pressures is crucial for understanding the dynamics and properties near the Earth's core. Traditional DFT methods fall short of accurately predicting the electronic structure of NiO and Ce due to strong correlation effects, underscoring the need for more advanced simulation techniques.

In response to these challenges, this study introduces a novel approach that leverages machine learning (ML) to develop interatomic potentials from quantum embedding methods, aiming to transform the study of strongly correlated systems. By combining these beyond-DFT techniques with the robust interpolation capabilities of graph neural networks, we propose a method to significantly reduce the time associated with simulating the structural properties of these complex materials while taking into account dynamical correlation effects. Specifically, we focus on the multi-orbital Gutzwiller approximation for $f$-electron materials and DFT+DMFT for $d$-electron materials as computationally tractable methods to generate the initial training data for our ML models. This approach enables us to capture the essential physics of correlated systems with distributed computational cost, offering a promising pathway to accurately simulate phase transitions in Ce and the melting behavior of NiO under extreme conditions.  Through the innovative use of machine learning interatomic potentials (MLIPs), we offer a scalable approach to explore and understand the complex structural dynamics of strongly correlated materials, contributing to the accelerated development of novel materials with desirable properties.

\section{Methods}

Local Density Approximation (LDA) calculations were carried out using the augmented plane wave plus local orbital method, as implemented in the WIEN2K package \cite{wien2kblaha2020}. For Ce, a muffin tin radius of 2.5 Bohr was utilized, while for Ni and O, the radii were set to 1.8 Bohr and 1.5 Bohr, respectively. The smaller muffin tin radii for the NiO system accommodated high-pressure calculations. RKMax was set to 8.5 for the Ce system and 7.5 for the NiO system. For the Cerium calculations, 500 k-points were used while the NiO calculations used 1000 k-points. LDA calculations were converged to a charge density within $10^{-3}$ and energy within $10^{-3}$ Rydberg. DMFT calculations for both systems were carried out using the eDMFT package \cite{dcHaule2010,edmftfHaule2015}, with the impurity problem solved through a Continuous Time Quantum Monte Carlo solver \cite{ctqmcHaule2007}. Forces were obtained as a derivative of the Luttinger-Ward functional with respect to atomic positions \cite{edmftfforcesHaule2016}.

For the Cerium system, charge self-consistent Gutzwiller calculations were carried out for every training and testing structure using the CyGutz package as implemented in \cite{Lanata2015,Lanata2017} until energy change was less than $5\times10^{-4}$ Rydberg. The Hubbard U parameter chosen was 6.0 eV with a Hund's coupling of 0.7 eV, since previous studies indicate these values provide good agreement with experimental lattice parameters \cite{hubbardUlanata2013,Lanata2014}.

Training structures were generated by producing 50 intermediate structures between the face-centered cubic and each of the 2 intermediate pressure stable phases. The intermediates were created by linearly interpolating the lattice vectors and the atomic positions. Those intermediates were then each perturbed 10 times, leading to 500 random structures generated for each pathway. The face-centered cubic phase was also perturbed 499 times leading to an additional 500 structures. Perturbations to a structure include displacement of the unit cell basis vector lengths by a random amount between 0.4 $\angstrom$ and -0.4 $\angstrom$, a change in the angles by a random amount between 20 to -20 degrees, and the atomic positions randomly displaced from their original positions with a displacement chosen from a normal distribution with a standard deviation of 0.15 $\angstrom$ and a mean value of 0. Each of the perturbations are performed one after the other in the order stated. Charge self-consistent LDA+Gutzwiller calculations were then carried out on each structure to obtain values for internal energy. 

The resulting energies were used to train a graph neural network using the M3GNET package \cite{m3gnetChen2022}. 1350 structures were randomly selected as training data, with the remaining 150 structures used for validation. Since force data is not yet easily obtainable within the Gutzwiller approximation, the finite difference method was used to calculate the forces and unit cell stresses after generation of the MLIP. 

The solid state nudged elastic band (NEB) method was then used to discover transition pathways between stable structures \cite{nebHenkelman2000}. This method involves fixing the starting and ending structures, attaching springs to the atoms and unit cell parameters of the intermediate structures, and optimizing the intermediate structures until a stable pathway is found. The MLIP was used as the mapping between the crystal structure and the energies, forces and stresses. The pathways found were checked using charge self-consistent DMFT. We checked the pathways at temperatures of 116 K and 400 K for the impurity solver with $1.28\times10^9$ Monte Carlo steps split across 16 processors. Results converged on average at 10 charge-self consistency steps defined as the point when variation of the energy dropped below $10^{-3}$ eV. The same values of Coulomb repulsion and Hund's coupling used for the Gutzwiller solver were used for the DMFT calculations. We employed nominal double counting as described in Refs. \cite{dcPourovskii2007,dcHaule2014,dcHaule2010} to correct for the double contribution to the energy from both the LDA and DMFT solvers with a nominal value of 1.0 for Ce. Nominal double counting has been shown to perform better than the fully localized limit method used in many other studies \cite{dcHaule2014}. 

For Nickel Oxide, full DMFT calculations were carried out to compute energies and forces. A Coulomb repulsion value of 8.0 eV and Hund coupling of 0.9 eV were chosen as they have been found to provide a good description of the electronic structure of NiO within DMFT at high pressures \cite{Gavriliuk2023}. A double counting value of 8.0 was chosen and calculations were carried out at a temperature of 611 K, which is above the Neel temperature of NiO. Approximately $8\times10^8$ Monte Carlo steps were sampled, split across 8 processors.

The training data was generated by performing 500 random perturbations of the atomic positions and lattice constants for the trigonal, face-centered cubic and body centered cubic phases each for a total of 1500 training structures. For each structure, the volume was also randomly perturbed, down to 65\% of the equilibrium volume, in order to capture behavior at high pressures. After the volume change, perturbations to the structure include displacement of the unit cell basis vector lengths by a random amount between 0.3 $\angstrom$ and -0.3 $\angstrom$, then a change in the angles by a random amount between 22.5 to -22.5 degrees, and displacement of the atoms in the unit cell from their original positions by a random value between -10\% to 10\% of each of the newly changed unit cell vectors. The same training and validation split of 1350 to 150 was used for NiO as for Ce. 

Molecular dynamics simulations were carried out using the atomic simulation environment package (ASE) \cite{aseLarsen2017} in the micro-canonical (NVE) ensemble to investigate the evolution of the melting curve under pressure using the Z-method \cite{zmethodBelonoshko2006,zmethodBelonoshko2008}. The NVE ensemble fixes particle number (N), volume (V) and conserves the internal energy (E). This is suitable for the Z-method as it involves heating a structure at a constant volume until it melts, and observing the pressure and temperature where the melt occurs.

Experimental volume-pressure relations were compared to DFT calculations from the VASP package \cite{vaspKresse1993,vaspKresse1996,vaspKresse1996_2} using a plane-wave basis set with Planar augmented wave pseudopotentials \cite{pawBlochl1994,pawKresse1994,pawKresse1999} and an exchange-correlation functional based on a revised Perdew-Burke-Ernzerhof for solids (PBESol) version of the generalized gradient approximation \cite{pbesolPerdew2008}. Additional calculations were also carried out using the standard Perdew-Burke-Ernzerhof (PBE) functional \cite{pbePerdew1996} when appropriate.

The M3GNET Graph neural networks take as input the atomic numbers, positions and lattice constants for a given unit cell of a crystal structure. The atomic positions are then used to derive edge bonds between neighboring positions, as well as three body interactions, such as bond angles. Atomic numbers are converted to a vector of dimension 64 using the built in graph featurizer. Edge vectors between 2 atoms were generated using basis functions proposed by Ref. \cite{m3gnetKocer2019}, with the distance between the atoms as input and a cutoff of 5 $\angstrom$. Three-body features were restricted to a cutoff radius of 4 $\angstrom$. These features were then fed into a graph convolution layer 4 times, with the embeddings being adjusted at each pass, before being passed to a gated multi-layer perceptron. The multi-layer perceptron then maps the graph convoluted input to energies, and can be differentiated to obtain forces and stresses. It consists of 3 layers with 64 weights for the first 2 hidden layers, and 1 output weight for the last layer. Further details on the construction of the interatomic potentials from graph features, as well as details of the model architecture can be found in Ref. \cite{m3gnetChen2022}. The same model architecture as described above was used for both Cerium and Nickel Oxide. The weights of the MLIP were adjusted using the AdamW optimizer \cite{m3gnetIlya2017} until the mean average error of the energies was less than 50 meV/atom for Cerium and 20 meV/atom for NiO. The weights of the neural network can also be refined by force data. In the case of Cerium, we have trained a single neural network to predict the energies by comparing the predicted energies by the model to the actual energies, then updating the weights based on the difference. For Nickel Oxide, we have trained two separate models, one to predict energy while the other predicts forces. This was achieved by reducing the adjustment of the weights based on the difference in predicted and actual energies to 1/10$^{th}$ of the weight the difference in predicted and actual forces in the force model, and similarly reducing the impact of the predicted to actual forces in the energy model. The force model was trained until errors were below 0.2 eV/$\angstrom$. 


\section{Results and discussion}

\begin{figure}

    \includegraphics[width=8cm]{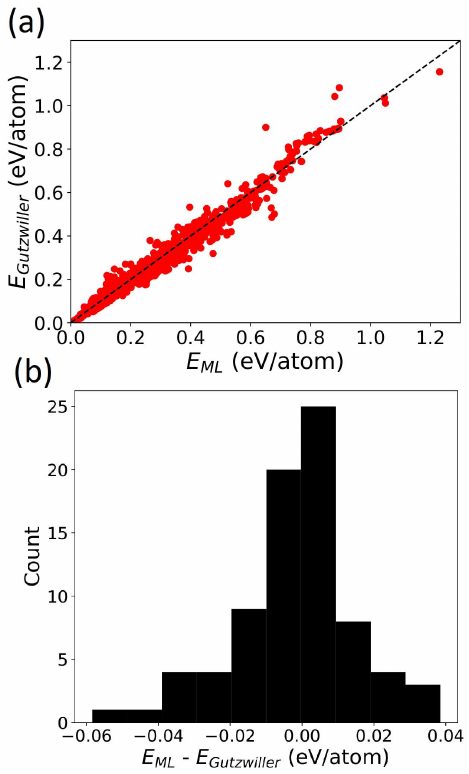}
    \caption{(a) Internal energies predicted by the MLIP versus energies calculated by solving tight binding model with Gutzwiller approximation for various Ce structures within the training set. Black dashed line represents perfect agreement. The points represent the initial 1500 Ce structures that were generated from random perturbations of cell vectors and atomic positions of both interpolated pathways and the $\alpha$ phase and (b) Histogram of difference between MLIP prediction of internal energy and Gutzwiller predicted internal energy for structures in the validation dataset whose DMFT energy is within 200 meV/atom of the energy of the $\alpha$ phase.}
    \label{fig:Ceallaccuracy}
\end{figure}

As a prototypical correlated $f$-electron material, elemental Ce has been extensively studied for decades \cite{ceKoskenmaki1978,ceKrisch2011,ceKim2019}, with computational focus directed at the isostructural $\gamma \rightarrow \alpha$ transition, which has been described as a collapse due the large decrease in volume ($\approx$ 17\%) as pressure increases past 0.8 GPa \cite{ceEryigit2022}. 
Efforts have also been made to explore the intermediate pressure phases of Ce within the 5-12 GPa range, where it typically undergoes a structural phase transition from a face-centered cubic phase ($Fm\overline{3}m$, $\alpha$-Ce) to either a monoclinic ($C2/m$, $\alpha''$-Ce) or orthorhombic ($Cmcm$, $\alpha'$-Ce) phase, depending on sample preparation conditions \cite{ceHuang2019,ceMunro2020,ceJohansson2014}. However, the stability and synthesis conditions for these intermediate-pressure pathways remain ambiguous. Some studies advocate for the monoclinic phase as most stable \cite{ceMcMahon1997,ceOlsen1985} while others argue that the orthorhombic phase \cite{ceZhao1997,ceGu1995} is most stable at these conditions. Recent investigations suggest that the orthorhombic phase is favored at higher temperatures, whereas cold-working tends to stabilize the monoclinic phase \cite{ceMunro2020}.  There is a notable scarcity of computational studies investigating the transition pathways for these phase transformations.
Traditional DFT struggles to account for the correlation effects present in Ce, and beyond DFT methods are required for accurate descriptions of the potential energy surface. In addressing the computational complexity associated with exploring these transition pathways, our study employs a MLIP, trained with data from LDA+Gutzwiller calculations, to investigate the transition pathways from the $\alpha$ phase to these intermediate pressure phases.

One limitation of the Gutzwiller solver is the neglect of electronic entropy. Since the Gutzwiller solver runs at T = 0 K,  it does not account for the mixing of higher energy states, and the ground state calculated is simply the lowest energy state. While this limitation is not expected to significantly affect the final pathway calculated, it is important to note that finite temperature extensions of the Gutzwiller solver \cite{gutzwillerLanata2015} may be employed in the future to increase the accuracy of results.

Predicted energies from our interatomic potential are in agreement with those obtained from the Gutzwiller solver, as demonstrated in our results (Fig. \ref{fig:Ceallaccuracy}(a)). While higher energy phases show more errors, the accuracy for low energy structures (critical for transition pathways) is quite high, which lends confidence to our computational predictions, as transition pathways are typically comprised of structures from the low energy regime. Shown in Fig. \ref{fig:Ceallaccuracy}(b) is a histogram of the errors for structures with a Gutzwiller energy within 200 meV/atom of the $\alpha$ phase. For these low energy structures, the error is clustered more tightly around 0 than for higher energy structures. The energies observed along the transition pathway are expected to stay within 100-150 meV/atom of the $\alpha$ phase, as typical barrier heights generally do not exceed this value \cite{ssnebYang2024, ssnebLyu2024}. The mean average error for all the structures is 50 meV/atom. However, for structures with Gutzwiller energies within 200 meV/atom of the $\alpha$ phase energy, the mean average error is 11.4 meV/atom.

Furthermore, since we are calculating the final free-energy values using a DMFT solver, the relative energies $\alpha$, $\alpha'$ and $\alpha''$ phases should be highly accurate and not susceptible to the error produced within the MLIP. That error will show itself within the transition barrier height as the MLIP may choose a less optimal transition pathway. However, the barrier height difference is greater than the 11.4 meV/atom error, and the free energies of the final transition pathway are calculated using full DMFT, which should provide accurate free energies independent of the method used to generate the structures along the transition pathway. With DMFT free energies typically showing an error of 1 meV, this corresponds to a temperature error of approximately 11.6 K, which is negligible compared to the temperature range investigated here.

\begin{figure}
    \includegraphics[scale=0.35]{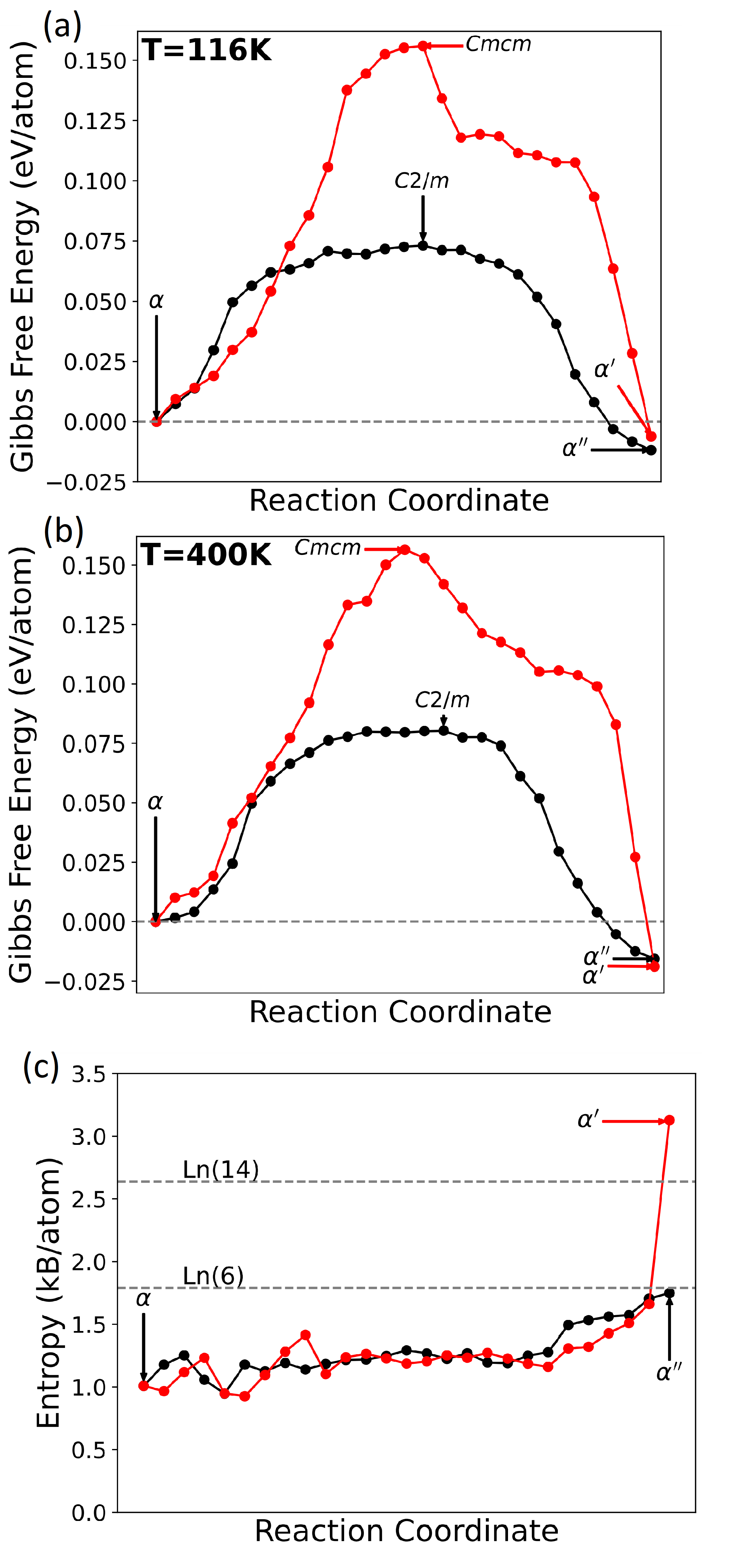}
    \caption{Gibbs Free Energy per atom for the $\alpha \rightarrow \alpha''$ (black) and $\alpha \rightarrow \alpha'$ (red) transition pathways at a pressure of 7.5 GPa, calculated using single-site DMFT at a temperature of 116 K (a) and 400 K (b). (c) Entropy per atom for the $\alpha \rightarrow \alpha''$ (top) and $\alpha \rightarrow \alpha'$ (bottom) transition pathways at 7.5 GPa of pressure calculated using single-site DMFT at a temperature of 400 K. Since spin-orbit coupling tends to split Ce orbitals into combinations of 6-fold and 8-fold degenerate states, the dashed lines indicate the full single-particle local space dimension entropy (Ln(14)) and the entropy of the 6-fold degenerate states (Ln(6)).}
    \label{fig:atoappCepathway}
\end{figure}

Using the solid-state NEB method, we explored transitions under 7.5 GPa of pressure and checked the results using a full DMFT treatment calculated at 116 K and 400 K. We calculated transition pathways between each of the two phases using the MLIP as a solver for the solid state NEB method. Since the end points are fixed, we do not expect to see overshoots in the beginning and final structures. As our goal is to investigate potential experimental transition pathways to the $\alpha'$ and $\alpha''$ phases from the $\alpha$ phase, overshoots are not necessary in this case. 

The initial and final structures were chosen to minimize the Gibbs Free Energy at 7.5 GPa. For the $\alpha$ phase, structural parameters were obtained from Ref. \cite{ceOlsen1985} with data taken from the measurements at a pressure of 0.94 GPa. For the $\alpha'$ and $\alpha''$ phases, structural parameters were taken from Ref. \cite{ceMcMahon1997} for the data at 7.5 GPa. For all phases, a Gibbs free energy versus volume graph was constructed and the volume that minimized the Gibbs free energy at 116K was chosen for the endpoints.

Since the MLIP is trained on structures near the phase space of linearly interpolated pathways between each of the stable phases, the transition pathway found will be comprised of structures close to the linearly interpolated pathway as well. It is unlikely that the transition pathway would take the structures very far away from this regime as the energetic cost of such a transition would be quite high. However, the possibility exists and can be further investigated with a more accurate generalized potential comprised of training data from regimes outside the scope of this study. Since the solid state NEB method discovers only one pathway for each phase transition, other techniques such as umbrella sampling \cite{altssnebKastner2011}, evolutionary algorithms \cite{altssnebOganov2006}, or swarm intelligence methods \cite{zhuPhaseTransitionPathway2019} may be employed in the future to increase the accuracy of results. Training the MLIP on finite temperature data may also lead to greater accuracy for specific temperatures.

The free energies of the structures along the predicted transition pathways were then calculated using DMFT at 116 K and 400 K to obtain the shifts in energetics at different temperatures. DMFT includes electronic entropic contributions at finite temperatures, making it more suitable for checking the energetics of the predicted pathways than the Gutzwiller method. In addition, using DMFT allows us to mitigate the errors introduced by the MLIP by providing a full re-calculation of the free energy at a finite temperature. However, generating training data for $f$-electron materials with many DMFT calculations is extremely computationally cumbersome. Therefore, the Gutzwiller solver is preferable for training data generation, as it captures many of the shifts in electronic density produced by strong correlation effects while allowing for rapid calculations.

At the lower temperature of 116 K (Fig. \ref{fig:atoappCepathway}a), the $\alpha$ phase is predicted to transition to the monoclinic $\alpha''$ phase, which has the lowest Gibbs free energy.  The transition to the orthorhombic $\alpha'$-Ce phase, while still exothermic, involves a higher reaction barrier and higher Gibbs free energy at 116 K, suggesting a less favored phase transition under these conditions. These findings agree with the experimental observations and support the hypothesis that cold-worked samples prefer the monoclinic phase, as it has the lowest Gibbs free energy at 116 K. The difference in the height of the reaction barriers is around 80 meV/atom, which is quite a bit higher than the error of the MLIP for structures within 100 meV of the $\alpha$ phase and larger than the energy scale of the temperatures under consideration. While the exact value may differ, the qualitative behavior of the $\alpha'$ pathway having a higher reaction barrier at this temperature should hold. The differences in free energies for the two phases should also be accurate as those are calculated using DMFT with no error introduced by the MLIP.

As the temperature increases to 400 K, the energy landscape shifts (Fig. \ref{fig:atoappCepathway}b). The reaction barriers stay at roughly the same height. However, the orthorhombic $\alpha'$-Ce phase becomes the lowest energy state by around 3 meV/atom, indicating an increased likelihood of obtaining $\alpha'$-Ce at higher temperatures.
This observation aligns with the established phase diagram and supports the hypothesis that the $\alpha''$ phase is a metastable state at higher temperatures \cite{ceMunro2020}. The reduction in observation of the $\alpha''$ phase at such temperatures can be attributed to the role of thermal energy in favoring the orthorhombic phase, even with a higher reaction barrier. 

It should be noted that the equilibrium structures may not appear to be saddle points in the pathway as the flattening is not complete near the endpoints. This is due to the different methods used to calculate the pathway. Specifically, the equilibrium phase volumes are optimized using the Gibbs free energy from DMFT, while the intermediate states are calculated using the solid-state NEB method with the MLIP. This causes the curvature near the endpoints to be slightly steeper than if one method was used throughout. The MLIP also predicted slightly lower barrier heights, around 20 meV lower for the $\alpha'$ phase and 10 meV lower for the $\alpha''$ phase compared to DMFT, although the general shape of the curve remained the same. 

We also track entropy across the transition pathway as shown in Fig. \ref{fig:atoappCepathway}c at a temperature of 400 K. Since spin-orbit coupling splits the Cerium $f$-orbitals into 6 degenerate $j=5/2$ and 8 degenerate $j=7/2$ states, we represent the entropy of the single particle local space dimension of the 6-fold degenerate $j=5/2$ states by $k_B Ln(6)$ line and the entropy of the full orbital by the $k_B Ln(14)$ line. Entropy calculations further elucidate the phase stability, revealing higher entropy for the $\alpha'$ phase, which corroborates its thermal stabilization.

We report only the 400 K run results for the entropy calculations as CTQMC cannot accurately sample the parts of the partition function required to reliably estimate electronic entropy as low temperatures. This is due to the sampling of terms in the hybridization expansion of the action being pushed to higher orders as temperature is lowered. Since the technique for estimating electronic entropy involves the zeroth order term, this leads to problems at low temperatures where insufficient sampling of the low order diagrams makes data unreliable \cite{edmftfHaule2015}.

A large coexistence region of these two phases is reasonable in light of the competing mechanisms. Based on our calculations, the entropic contributions are strong for the $\alpha'$ phase, which competes against the lower reaction barrier for the $\alpha''$ pathway. Given that fluctuations in temperature during sample preparation are inevitable, either phase may be produced depending very heavily on the conditions of preparation.
It is also worth noting that reaction barriers may be substantially larger when elastic effects are taken into account, as indicated by Bustingorry et al. \cite{bustingorry2005thermodynamics}. In our current study, the reaction barriers are calculated without explicit elastic interactions between domains or grains, which may lead to an underestimation of the true barrier height. Accounting for elastic effects in future calculations could provide a more comprehensive view of the phase stability and transition pathways, particularly under non-uniform stress conditions.

\begin{figure*}
    \includegraphics[width=17cm]{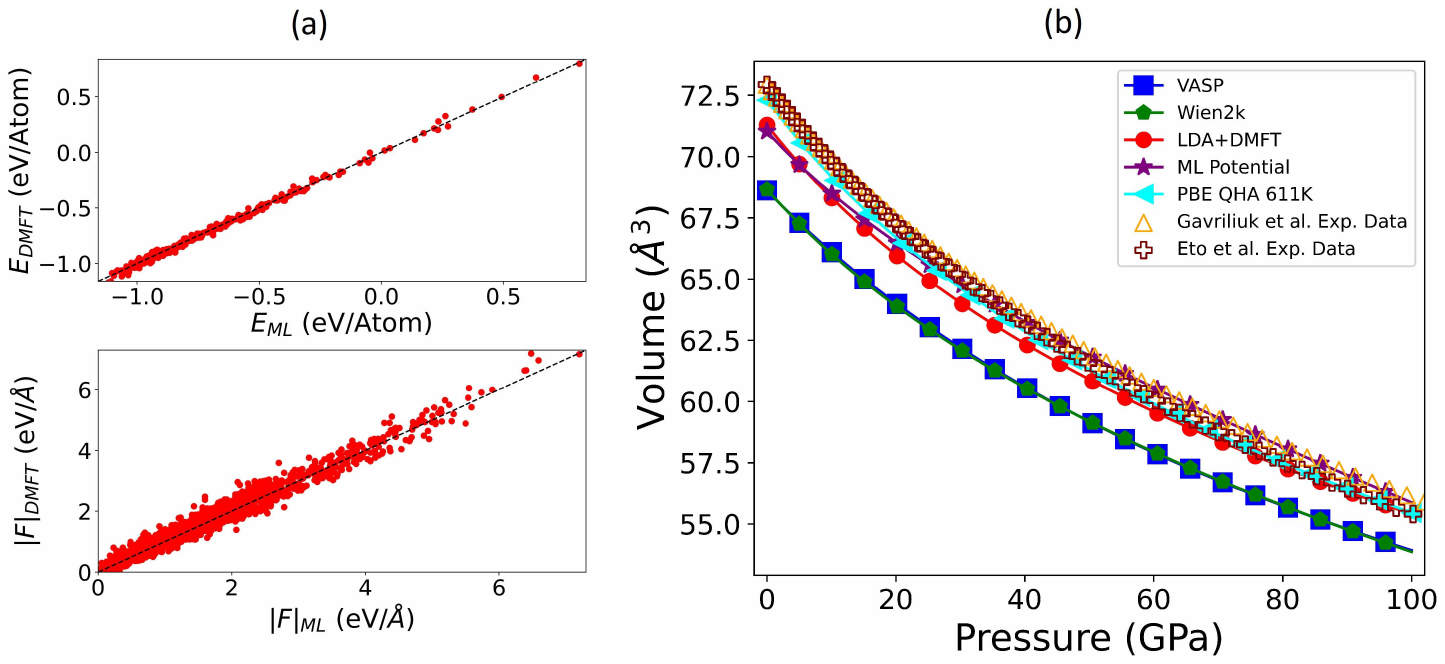}
    \caption{(a) Internal energies and forces predicted by the MLIP versus energies and forces from LDA+DMFT for various NiO structures within the training set. Black dashed line represents perfect agreement. The dots represent each of the 1500 structures that were used to generate the MLIP. Energy has been normalized by subtracting the average energy in the set of DMFT energies before training. The mean average error on the validation set was 19 meV/atom. (b) Comparison of equilibrium volumes of FCC phase of NiO under pressure from various first-principles programs to machine learning prediction and to experimental volume relations generated from the Birch–Murnaghan equation of state as reference. LDA+DMFT calculations were run at a temperature of 611K. Yellow hollow triangles were obtained by fitting parameters from Ref. \cite{Gavriliuk2023} while the brown hollow crosses were obtained by fitting parameters from Ref. \cite{nioEto2000}.}
    \label{fig:nioaccuracy}
\end{figure*}

Building upon our insights from elemental Ce, we extend our investigation to NiO, a prototypical Mott insulator emblematic of strongly correlated compounds. First studied in the 1930s \cite{niodeBoer1937,nioMott1937,nioMott1949}, the unique properties of NiO are the result of the strong correlation among its 3$d$ electrons \cite{nioGavriliuk2012,nioAnisimov1990}. Experimentally, it has been established that NiO remains insulating under extremely high pressures \cite{Gavriliuk2023}, and NiO can be considered as a representative compound for the behavior of correlated materials under the extreme conditions found within the Earth. Accurate predictions of large-scale thermal properties, such as melting curves, are essential for understanding the structure inside the Earth. These properties, however, are challenging to calculate with conventional DFT due to its failure to accurately describe the ground state of strongly correlated materials \cite{nioDelloStritto2023,nioGebhardt2023}.

Addressing these challenges, we use graph neural networks trained on energies and forces from LDA+DMFT calculations. DMFT inherently includes the dynamical correlation effects in the many-body Hamiltonian, offering an accurate prediction of electronic properties. DMFT is also less sensitive to the particular Hubbard U value used. Under pressure, the U value is liable to change, but this effect affects final results less when performing DMFT calculations as compared to techniques like DFT+U. This approach also benefits from the inclusion of many-body effects in force calculations, potentially leading to significant structural insights when contrasted with standard DFT predictions \cite{cePlekhanov2018}.
To ensure sufficient transferability, the MLIP is trained
on a variety of data generated by random structure generation as well as energy versus volume data generated from the face centered cubic ($Fm\overline{3}m$), trigonal ($R\overline{3}m$), and body centered cubic ($Pm\overline{3}m$) structures. LDA+DMFT calculations were run at 611 K, above the Neel temperature for NiO. Since our goal is to investigate melting curves, NiO will be within the paramagnetic phase near the melting temperature and the training data should reflect the target electronic structures. Higher temperatures for the DMFT solver may be chosen, but this does not affect the electronic structure significantly. Calculations done at 611 K and 2000 K using DMFT for the FCC structure show only a 0.004 difference in the occupation number of the impurity site. The full Coulomb interaction was taken into account when solving the impurity problem. 

The accuracy of the machine learning network is depicted in Fig. \ref{fig:nioaccuracy}a, showing close alignment between the predicted and actual energies and forces. The error for the 150 validation structures is around 19 meV/atom, which is far below the energy scale for the temperatures of interest. Since we are looking at temperatures of 2000 K and greater, this corresponds to an energy scale of at least 172 meV. The error of 19 meV/atom is fairly uniform across the whole energy scale, increasing slightly at energies more than 1 eV/atom above the lowest energy structure.

Figure \ref{fig:nioaccuracy}b shows the equilibrium volume predictions under varying pressures for the face-centered cubic phase of NiO, as derived from various computational methods and techniques. The LDA+DMFT calculations were run at a temperature of 611 K so that NiO is within the paramagnetic phase. Wien2k \cite{wien2kblaha2020} and VASP \cite{wien2kblaha2020} were both run with the PBESol \cite{pbesolPerdew2008} GGA exchange correlation functional at zero temperature. Experimental results were gathered at room temperature, which is around 300 K. The MLIP shows good agreement with LDA+DMFT calculations in a wide pressure range, indicating a high degree of accuracy. Notably, both the MLIP and LDA+DMFT calculations surpass the traditional DFT predictions from Wien2k and VASP when compared to the experimental data. The MLIP provides an energy versus volume curve in good agreement with that from LDA+DMFT, yet it achieves this at a significantly reduced computational cost.

We have also performed Quasi-Harmonic Approximation (QHA) calculations using the PBE functional within VASP as well as the Phonopy package \cite{phonopyTogo2010,phonopyTogo2023_2,phonopyTogo2023} in order to investigate thermal expansion effects. The QHA allows the inclusion of lattice vibration entropy as well as energy contributions from lattice vibrations in order to investigate the Gibbs free energy under varying temperature and pressure. The teal left-pointing arrows in Figure \ref{fig:nioaccuracy}(b) show the equilibrium volume at a temperature of 611 K for QHA calculations run using the PBE GGA functional  \cite{pbePerdew1996}. The PBESol functional was not used as it indicated that volume decreases at high temperatures. A previous study has found PBE to be closer to experimental lattice constants in AFMII NiO, which PBESol significantly underestimates \cite{nioDelloStritto2023}. We have also found PBESol to underestimate lattice constants in the paramagnetic cubic phase. The temperature of 611 K was chosen in order to give a fair comparison with the ML potential and the LDA+DMFT calculations. The agreement with experiment is quite good for the QHA calculations and show that lattice vibrational entropy is quite significant in NiO. However, since zero temperature DFT calculations are used for these calculations, results may be quite different if the ground state charge density at zero temperature differs significantly from the charge density at finite temperatures. DMFT should provide more consistent results at finite temperature across a wider range of structures.

Next, we performed molecular dynamics calculations to estimate the melting curve of NiO under high pressures. Theoretical studies on the melting curves of strongly correlated materials are rare, due to the challenges associated with accurately simulating these phenomena across different pressure conditions. Leveraging a neural network that directly predicts energies and forces enables us to perform supercell calculations more cost-effectively than traditional DFT, while still capturing some of the force renormalization effects. 

\begin{figure}
    \includegraphics[width=10cm]{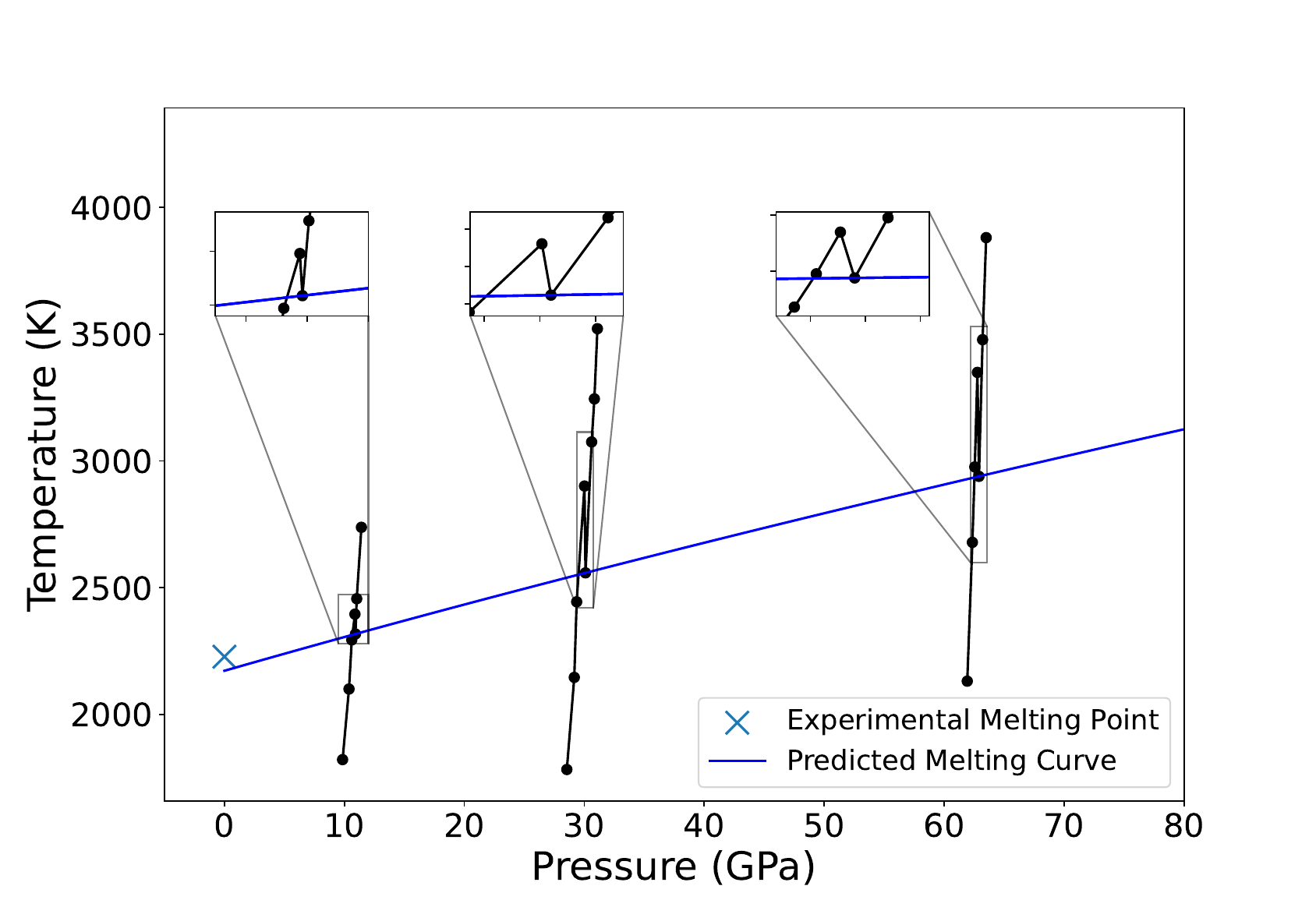}
    \caption{Melting curve in Pressure-Temperature space. Shown in black are the Z-method curves for 3 isochores at 70 $\angstrom^3$, 65 $\angstrom^3$, and 60 $\angstrom^3$ in order of increasing pressure. The curves show an increase in temperature until the crystal reaches a superheated state, after which the pressure increases with a simultaneous drop in temperature, as shown in the insets. The melting curve is fit the point right after the melt occurs.}
    \label{fig:70angzmethod}
\end{figure}

For our melting curve investigation, we used the Z method owing to its efficiency and proven reliability in experimental comparisons \cite{zmethodBelonoshko2012,zmethodMoriarty2012,zmethodBouchet2009,zmethodKaravaev2016}. Focusing on high-temperature conditions, we constructed $3\times3\times3$ supercells derived from the face-centered cubic structure for our molecular dynamics simulations in the NVE ensemble with a spread of initial temperatures from a Maxwell-Boltzmann distribution. 
Our simulations were performed near the melting point, which is well above the Neel temperature of 525 K. Therefore, the starting structure for the molecular dynamics simulation was chosen to be the face-centered cubic structure ($Fm\overline{3}m$), as this is the structure that NiO adopts above its Neel temperature \cite{nioMandziak2019,nioEto2000}.
The system was allowed to equilibrate over 10 picoseconds with a timestep of 1 femtosecond, with thermal averages of relevant observables taken over the final 2 picoseconds. The melting curve was modeled using the Simon-Glatzel equation, a method that has been successful with other transition metal compounds \cite{nioErrandonea2013,nioHrubiak2017,nioParisiades2019,nioDewaele2007}, represented as:
\begin{equation*}
    T_m = 2172.42(1+\frac{P}{95.85})^{0.5993},
\end{equation*}
where $T_m$ is the melting temperature at a pressure $P$. The parameters of the Simon-Glatzel equation were obtained by a non-linear least squares fit to the pressure-temperature points along the isochores where the full melt occurred. This is indicated by an increase in pressure and a simultaneous drop in temperature.
Fig. \ref{fig:70angzmethod} presents the melting curve of NiO in Pressure-Temperature (PT) space, mapped using isochores of 70 $\angstrom^3$, 65 $\angstrom^3$, and 60 $\angstrom^3$. The MLIP effectively captures the melting behavior of NiO, with the predicted melting curve aligning well with experimental data at atmospheric pressure \cite{nioCRCMeltpoint}. The alignment of the MLIP with experimental data suggests a reliable model for predicting the behavior of strongly correlated materials under extreme conditions akin to those deep within the Earth. Therefore, this approach, integrating data from DMFT into graph neural networks, offers a novel and efficient pathway to explore the complex melting behavior of strongly correlated materials like NiO.

\section{conclusion} 
In summary, we have demonstrated the utility of MLIPs in accelerating investigations of dynamics of correlated systems. For Cerium, training data for the MLIP was generated using the LDA+Gutzwiller method, as generation of force data from DMFT for $f$-electron compounds with significant spin-orbit coupling is beyond our computational capabilities. The finite difference method was used to calculate forces and stresses, and these in turn were used to predict viable transition pathways between intermediate pressure phases of Cerium. The resulting transition pathway supports the claim that the $\alpha''$ phase is more stable than $\alpha'$-Ce at low temperatures. The transition barrier to the $\alpha'$ phase decreases with increasing temperature while the Gibbs free energy of the $\alpha'$ phase decreases below that of the $\alpha''$ phase at high temperatures. Given the larger reaction barrier to the $\alpha'$ phase at both low and high temperatures, but lower Gibbs energy at high temperatures, it is no surprise a large coexistence region for these two phases exists. For NiO, full DMFT calculations became feasible due to the smaller orbital size and the reduced effects of spin-orbit coupling. While still quite expensive to converge, training data of both energies and forces were obtained and an ML interatomic potential was trained on this data. Molecular dynamics simulations were carried out to determine the melting point of this material, with good agreement to the experimental ambient-pressure melting temperature. We hope in the future that more MLIPs can be trained for correlated systems, unlocking new avenues of transition state searching, theoretical thermodynamic predictions, and eventually crystal structure prediction of strongly correlated systems.  

\begin{acknowledgements}
This work was supported by the startup funds of the office of the Dean of SASN of Rutgers University-Newark. The authors acknowledge the Office of Advanced Research Computing (OARC) at Rutgers for providing access to the Amarel cluster and associated research computing resources.
\end{acknowledgements}

\bibliography{biblio}

\end{document}